\documentclass[journal]{IEEEtran}
\usepackage{cite}
\usepackage{amsfonts}
\usepackage{setspace}

\usepackage{setspace}
\usepackage{amssymb}
\usepackage{epsfig}
\usepackage{graphicx}
\usepackage{xcolor}
\usepackage{algorithm}
\usepackage{algorithmicx}
\usepackage{algpseudocode}
\usepackage{amsmath}

\floatname{algorithm}{Algorithm}

\usepackage[ps2pdf,
bookmarks=false,
bookmarksnumbered=false, 
bookmarksopen=false, 
colorlinks=false]{}

\begin{document}

\title{Adversarial Jamming Attacks and Defense Strategies via Adaptive Deep Reinforcement Learning}

\author{Feng Wang, Chen Zhong, M. Cenk Gursoy, and Senem Velipasalar
\thanks{The authors are with the Department of Electrical
		Engineering and Computer Science, Syracuse University, Syracuse, NY, 13244 (e-mail: fwang26@syr.edu, czhong03@syr.edu, mcgursoy@syr.edu, svelipas@syr.edu).}
\thanks{The material in this paper was presented in part at the 2020 IEEE Wireless Communications and Networking Conference (WCNC) and 2020 Annual Conference on Information Sciences and Systems (CISS) at Princeton University.}}

\maketitle

\begin{abstract}
As the applications of deep reinforcement learning (DRL) in wireless communications grow, sensitivity of DRL based wireless communication strategies against adversarial attacks has started to draw increasing attention. In order to address such sensitivity and alleviate the resulting security concerns, we in this paper consider a victim user that performs DRL-based dynamic channel access, and an attacker that executes DRL-based jamming attacks to disrupt the victim. Hence, both the victim and attacker are DRL agents and can interact with each other, retrain their models, and adapt to opponents' policies. In this setting, we initially develop an adversarial jamming attack policy that aims at minimizing the accuracy of victim's decision making on dynamic channel access. Subsequently, we devise defense strategies against such an attacker, and propose three defense strategies, namely diversified defense with proportional-integral-derivative (PID) control, diversified defense with an imitation attacker, and defense via orthogonal policies. We design these strategies to maximize the attacked victim’s accuracy and evaluate their performances.

\end{abstract}


\thispagestyle{empty}

\section{Introduction}
There has recently been growing interest in employing machine learning to address certain challenges in communication systems, such as modulation classification \cite{8645696}, user association and resource allocation \cite{8796358}, power control for interference management \cite{8855869} and dynamic multichannel access \cite{8303773}. In particular, dynamic access is a typical long-term control problem that is generally modeled as a Markov decision process (MDP) or a partially observable Markov decision process (POMDP). In addressing these problems, several recent studies considered the application of deep reinforcement learning (DRL), taking advantage of its ability to explore, learn and adapt in unknown environments. For instance, the authors of \cite{8303773} proposed a deep Q-network (DQN) to learn the channel switching pattern and make channel selection decisions. In \cite{8665952} and \cite{8532121}, DQNs are used to allocate the spectral resources in time-division multiple access and orthogonal frequency-division multiple access networks, respectively, with the aims to effectively utilize these resources in a multi-user setting. Instead of only learning the patterns in the environment, DRL algorithms can also be used to account for the intrinsic temporal correlations among sequential decisions. Such a DQN framework is proposed in \cite{8302493} with the goal to utilize the spectral resources and reduce the co-channel interference in multibeam satellite systems. Moreover, DRL algorithms are also applied to jointly solve the multichannel access and task offloading in mobile edge computing as presented in \cite{9037194}. Aside from the frequently used DQN, recent research has also explored the application of other DRL algorithms in solving these types of dynamic control problems. For instance, authors in \cite{9037194} made use of the actor-critic framework, while a multi-agent deep stochastic policy gradient framework was proposed in \cite{kassab2020multi} for dynamic spectrum access problems.

While significant strides have been made in solving the aforementioned problems, this increasing interest in applying machine learning algorithms to communication systems also brings forth potential security risks due to adversarial attacks. Since machine learning methods are highly data-driven algorithms, even a minor modification in the observation data can lead to dramatic changes in the learning-based decision policies \cite{goodfellow2014explaining}. Therefore, adversarial machine learning has been intensively studied to better understand the vulnerabilities of machine learning methods. 

In particular, in the literature, adversarial attacks have been considered and widely applied to deep learning-based classification problems, such as the classification of images \cite{dong2018boosting} \cite{dong2019evading} \cite{jia2019enhancing}, time series \cite{fawaz2019adversarial} and sound events \cite{subramanian2019adversarial}. In these cases, the victim models are trained and fixed, and the input data is accessible to the attacker, so that the attack can be realized by crafting adversarial examples to mislead the victim's decisions. This idea is also used in the attack on reinforcement learning-based tasks \cite{zhao2019blackbox} and \cite{huang2017adversarial}. However, in certain control problems, the observations of the reinforcement learning agents are not available to the attacker, making it infeasible to craft any adversarial examples. To tackle this difficulty, in \cite{gleave2019adversarial}, the authors trained a reinforcement learning-based adversarial policy instead. It is proved that adversarial jamming attack with either white-box or black-box setting may also lead to significant performance degradation \cite{yang2020enhanced}. Regarding adversarial attacks in wireless systems, a deep learning-based wireless jamming attack has been studied in \cite{shi2018spectrum} and \cite{erpek2018deep}, in both of which, the system consists of a single transmitter, a receiver, one background traffic source whose activity decides the channel state and a deep learning-based jammer.

We finally note that in addition to adversarial attacks, there is a growing body of work introducing various defense strategies against adversarial attacks on deep learning schemes (e.g., using convolutional neural networks \cite{madry2017towards,zhang2019theoretically,xu2017feature} and deep reinforcement learning \cite{zhan2020preventing}). Adversarial jamming attacks also stimulate further investigation on corresponding mitigating strategies against such attacks \cite{sagduyu2020wireless}.


Motivated by these considerations, we in this paper address a scenario in which a well-trained victim user performs DRL-based dynamic channel access \cite{8896945}. We first design a DRL based jamming attacker that employs a dynamic policy and aims at minimizing victim's channel access accuracy.  Both the victim user and jamming attacker can learn effective strategies with feedback from each other over time. Therefore, in this setting, we address a dynamic control problem \cite{zhong2020adversarial}.
Subsequently, we develop defense strategies for the victim user. In particular, we propose diversified defense with proportional-integral-derivative (PID) control, diversified defense with imitation attacker, and defense via orthogonal policies to mitigate the jamming attack and maximize victim's accuracy of accessing channels with favorable conditions \cite{wang2020defense}.

To the best of our knowledge, we for the first time in this paper analyze the interactions between a DRL-based dynamic channel access agent and a DRL-based jamming attacker agent. Our key contributions can summarized as follows:
\begin{itemize}
	
\item We design a DRL attacker agent that can observes the environment partially (e.g., one channel at a time), learns the channel access patterns of the victim user, and performs jamming attacks. We also propose a novel stop-retrain-attack (SRA) procedure for the attacker to observe and attack the victim efficiently.
	\item We consider a practical setting in which both the victim and attacker start with black-box settings, where they have no knowledge of each other's model structure (except in the diversified imitation defense strategy) or parameters. We study their interactions with each other. After interacting, both retrain their models to adapt to each other over time. 

\item We propose diversified defense and orthogonal policies defense strategies to protect the victim from being learned by the attacker. We evaluate the performances of the defense strategies and demonstrate significant improvements in the victim user's accuracy in dynamic channel access.
\end{itemize}

The remainder of the paper is organized as follows. In Section \ref{Sec: pre}, we describe the DRL based multichannel access victim user model. In Section \ref{sec: RL}, we introduce the proposed DRL based jamming attacker, evaluate its performance and also address the impact of power budget limitations. In Sections \ref{sec: pid} through \ref{sec: 2policies}, we develop and analyze three different defense strategies: 1) diversified PID defense; 2) diversified imitation defense; 3) defense via orthogonal policies. Finally, we provide experiment results and address attack detection in Section \ref{sec: exp} and conclude in Section \ref{sec: con}.

\section{Dynamic Channel Access Policies of the Victim User}\label{Sec: pre}
In this section, we introduce the background on DRL based dynamic multichannel access. As noted above, we consider an actor-critic DRL agent proposed in \cite{8896945} as the victim user to be attacked.
\subsection{Channel Switching Pattern}\label{subsec: channel}

\begin{figure}
	\centering
	\includegraphics[width=1.1\linewidth]{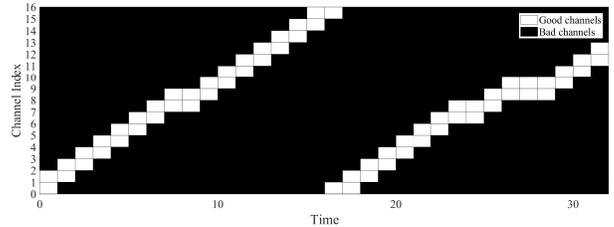}
	\caption{ Round-robin switching pattern when two of the 16 channels is in good condition and the switching probability is $\rho$ = 0.95. The channel in good state at a given time is indicated by white squares.}
	\label{fig:channelpattern}
\end{figure}

In the considered dynamic multichannel access problem, the time is slotted and the user selects one channel to access at the beginning of each time slot. We assume that the state of each channel switches between good and bad in a certain probabilistic pattern. When the channel is in good condition, the user can transmit data successfully. Otherwise, a transmission failure will occur. We also assume that the channel switching pattern can be modeled as a Markov chain, and in each state of which, there are $k$ out of the $N$ channels in good condition. At the beginning of each time slot, the channel pattern can either switch to the next state with probability of $\rho$, or remain to be the same as the state in the last time slot with probability of ($1-\rho$). In Fig. \ref{fig:channelpattern}, we display a round-robin switching pattern with two out of $16$ channels being good in each time slot and each channel has the same probability to be in good state.

\subsection{Actor-Critic Agent}\label{subsec: victim}

It is assumed that the channel switching pattern is unknown to the user, and the user can only observe the channel selected in the current time slot. Hence, the multichannel access is a partially observable Markov decision process (POMDP). To help the user to access the good channels as frequently as possible under such conditions, we proposed in \cite{8896945} an actor-critic deep reinforcement learning based agent to make the channel access decisions in each time slot.

The proposed agent is designed to learn the channel switching pattern through past decisions and the corresponding feedback from the channels. We assume that, at time $t$, the channel state can be denoted as $\mathtt{X}_t = \{\mathtt{x}_1, \mathtt{x}_2, \ldots, \mathtt{x}_{N} \} $, where $N $ is the total number of channels, $\mathtt{x}_i$ stands for the state of the $i^{th}$ channel. For each channel $i$, where $i = 1, 2, \ldots, N$, we have $\mathtt{x}_i = 1$ if the channel is in good state, or $\mathtt{x}_i = 0$ if the channel is in bad state. And each time the agent senses a channel, the state of the sensed channel is revealed to be either good or bad. Therefore, we define the reward (feedback) as follows: if a good channel is chosen, the reward $r_t$ will be $+1$; otherwise, the reward $r_t$ will be $-1$.

The agent's observation can be denoted as $O_t = \{o_1, o_2, \ldots, o_N \}$, where $N$ is the total number of channels. If channel $i$, $i = 1, 2, \ldots, N$, is chosen, the agent senses it and learns its state, so we define $o_i = r_t$; otherwise, the agent will record $o_i = 0$. The agent will learn on the basis of its previous experience. We assume that the agent keeps an observation space $\mathcal{O}$ that consists of the most recent $M$ observations. The observation space is initialized as an all-zero $N \times M$ matrix, and at each time $t$, the latest observation $O_t$ will be added to the observation space, and the oldest observation $O_{t-M}$ will be removed. The updated observation space $\mathcal{O}$ at time $t+1$ can be denoted as $\mathcal{O}_{t+1} = \{O_{t}; O_{t-1}; \ldots; O_{t-(M-1)} \}$.

Next, we consider a discrete action space denoted by $\mathcal{A} = \{1, 2, \ldots, N \}$, where $N$ is the total number of channels. Each valid action in the action space describes the index of the channel that will be accessed. Hence, when an action is chosen, the agent will access the corresponding channel and receive the reward which reveals the condition of the chosen channel. The agent can only choose one channel to sense/learn in each iteration. The aim of the agent is to find a policy $\pi$, which maps the observation space $\mathcal{O}$ to the action space $\mathcal{A}$,  that maximizes the long-term expected reward $R$ of channel access decisions:
\begin{equation*}
\pi^* = \arg \max_{\pi} R
\end{equation*}
where $\pi^*$ denotes the optimal decision policy, and in a finite time duration $T$, we express $R$ as
\begin{equation*}
R = \frac{1}{T} \sum_{t = 1}^{T} r_t.
\end{equation*}
And according to the definition of $R$, we have $R \in [-1, 1]$.

\subsection{Performance in the Absence of Jamming Attacks}
\begin{figure}
	\centering
	\includegraphics[width=0.9\linewidth]{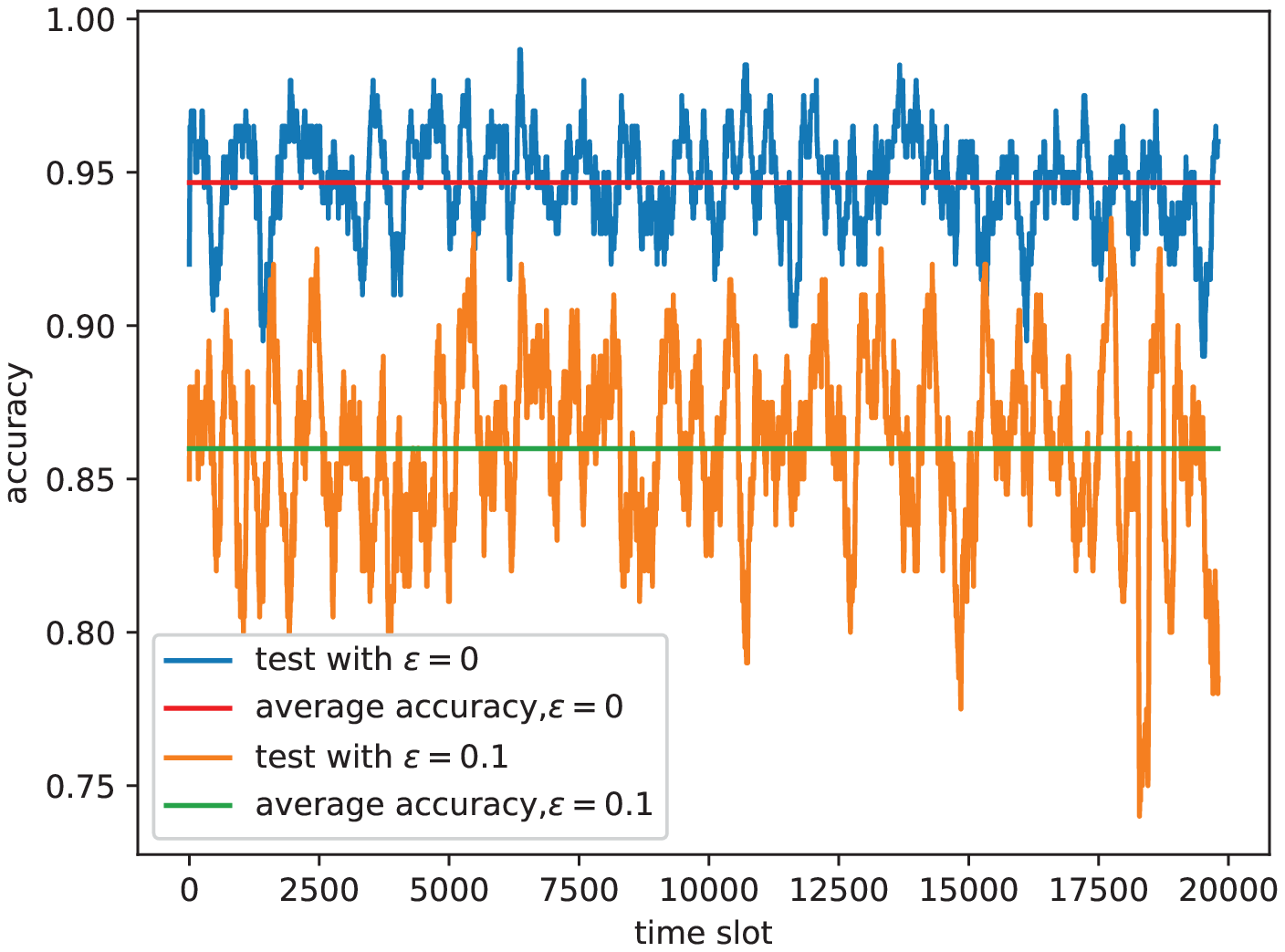}
	\caption{Accuracy of the good channel access in the absence of jamming attacks.}
	\label{fig:victimaccuracybeforeattack}
\end{figure}

We consider the channel switching pattern shown in Fig. \ref{fig:channelpattern}, and evaluate the accuracy of the good channel access by the user with $N = M = 16$. The evaluation is performed in the absence of any jamming attacks and after the DRL agent is well trained. In Fig. \ref{fig:victimaccuracybeforeattack}, we test the model in two cases. First, we consider the $\epsilon$-greedy policy with $\epsilon = 0.1$, with which the user accesses a random channel with probability 0.1 (for exploration), and chooses the channel selected by the reinforcement learning policy with probability 0.9. We note that $\epsilon$-greedy policies with $\epsilon > 0$ are generally employed to enhance the DRL agent's ability to adapt to changes in the channel patterns, as will be discussed in detail in Section \ref{sec: exp}. In addition, we also consider the case in which $\epsilon$ is set to 0 to identify the performance of the pure DRL policy. We observe in the figure that high average accuracies (higher than $85\%$ and around $95\%$ with $\epsilon = 0.1$ and $\epsilon = 0$, respectively) are attained in the absence of jamming attacks.

\section{DRL Based Jamming Attacker}\label{sec: RL}
In this section, we introduce an actor-critic DRL agent to perform the jamming attack on the aforementioned victim user without having any prior information about the channel switching pattern or the victim's action policy. The DRL attacker is able to jam a single channel in each time slot to significantly reduce the selection accuracy of the actor-critic agent. We also assume that the DRL attacker is able to observe the victim's interaction with the environment for a period of time that is sufficiently long for the DRL attacker to learn the activity pattern.

\subsection{Actor-Critic Model}

\begin{figure}
	\centering
	\includegraphics[width=.8\linewidth]{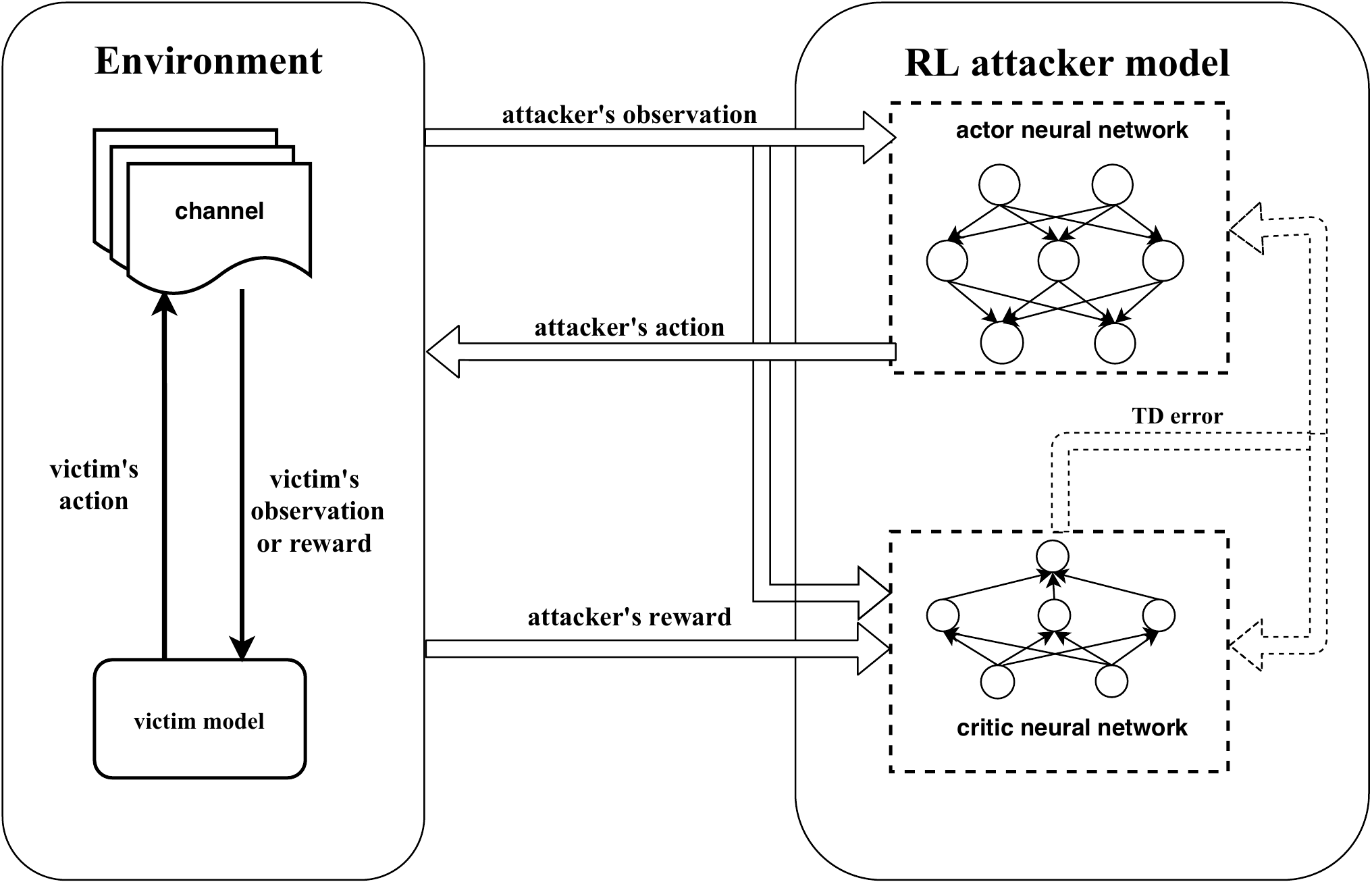}
	\caption{Diagram of actor-critic structure and DRL attacker-environment interactions.}
	\label{fig:rlattackerdiagram}
\end{figure}

In Fig. \ref{fig:rlattackerdiagram}, we show the diagram of the actor-critic structure and the DRL attacker-environment interactions. The channels and victim's channel selection model form the environment to be observed by the attacker. At the beginning of each time slot, after the DRL attacker observes the environment, it can select one channel based on its own observation and action policy learned by the actor-critic neural networks. Here, it is noted that the DRL victim also dynamically selects channels at the beginning of each time slot based on the victim's observations and action policy. We assume that the attacker and victim select channels without knowing the decision made by each other in the current time slot. Then, the attacker's reward and the new state of the environment after executing the chosen action will be sent to its critic neural network to calculate the temporal difference (TD) error. This TD error will be used to update both critic and actor neural networks. When the update of network is completed, the DRL attacker model is ready to make the next decision. Above, we have summarized the operation of the attacker DRL agent. Next, we provide more detailed descriptions regarding the observations, actions, rewards of this attacker (or equivalently jamming) agent along with its actor-critic structure.

\emph{Attacker Agent's Observations:} As mentioned before, the DRL attacker agent has no knowledge of the channel patterns and the victim user's policy.  Hence, from the perspective of the DRL attacker agent, the channels and victim form an unknown environment. And therefore, the only accessible information that can be used as the states in the actor-critic neural networks is the attacker agent's observations. We assume that in each time slot $t$, the observation of the DRL attacker is denoted as $\mathit{\Phi}_t = \{\phi_{t,1}, \phi_{t,2}, \dots, \phi_{t,N} \}$. Then each element $\phi_{t,i}$, for $i = 1, 2, \dots, N$,  stands for the observation in the $i^{\text{th}}$ channel at time $t$. As assumed before, the DRL attacker can only choose one channel at a time, so we have
\begin{align}
\phi_{t, i}=\begin{cases}
r_t \hspace{.5cm} \text{if the $i^{\text{th}}$ channel is selected in time slot $t$,}\\
0 \hspace{.5cm} \text{if the $i^{\text{th}}$ channel is not selected in time slot $t$.}
\end{cases}
\end{align}
Above, $0$ indicates that no information is available on a channel that has not been selected. And in our implementation, it is assumed that the DRL attacker can keep a memory of the latest $T$ observations, so that the memory forms the observation space $\mathbf{\Phi}_{t} = \{\mathit{\Phi}_{t-1}, \mathit{\Phi}_{t-2}, \dots, \mathit{\Phi}_{t-T}\}$ and use it as the input of the DRL agent.

\emph{Attacker Agent's Actions:} The action space of the attacker agent is formed by all valid actions that indicate which channel to jam to lead to victim's failure. Here, the attacker and victim share the same action space $\mathcal{A}$.

\emph{Attacker Agent's Reward:} The reward is received when the action is executed, meaning that the attacker agent chooses a channel to jam and gets direct feedback from the environment. The action of the DRL attacker and the victim at time $t$ are denoted as $a_t^A$ and $a_t^V$ respectively. Since both the DRL attacker and the victim select one out of the $N$ channels, the sizes of their action spaces are the same. We assume that there are proper mechanisms and measurements (such as SINR levels, ACK signals) through which the attacker learns if the victim has selected the same channel as the attacker itself, i.e., $a_t^A = a_t^V$, and if the victim has transmitted successfully. The goal of the attacking DRL agent is to learn the victim's activity pattern so that it can jam the channels selected by the victim as much as possible. Based on this objective, we define the reward of the DRL attacker at time $t$ as
\begin{align}
r_t = \begin{cases}
+ 1 \hspace{.55cm} \text{if $a_t^A = a_t^V$ and victim selects a good channel,}\\
+ 0.5 \hspace{.3cm} \text{if $a_t^A = a_t^V$ and victim selects a bad channel,}\\
- 0.5 \hspace{.3cm} \text{if $a_t^A \neq a_t^V$ and victim selects a bad channel,}\\
- 1 \hspace{.55cm} \text{if $a_t^A \neq a_t^V	$ and victim selects a good channel.}
\end{cases}
\end{align}
Within this setting, the DRL agent is encouraged to select the same good channels as the victim as its first priority. We also consider the case in which the attacker and victim select the same bad channel as partial success in terms of jamming.

\subsubsection{Attacker Agent's Actor-Critic Structure}  The actor-critic structure consists of two neural networks, namely the actor network and critic network. The parameters of these two neural networks are initialized and updated as follows:

\emph{Actor:} The actor is employed to explore a policy $\pi$, that maps the agent's observation $\mathbf{\Phi}$ to the action space $\mathcal{A}$:
\begin{equation}
\pi_{\vartheta}(\mathbf{\Phi}) : \mathbf{\Phi} \rightarrow \mathcal{A}.
\end{equation}
Therefore, the mapping policy $\pi_{\vartheta}(\mathbf{\Phi})$ is a function of the observation $\mathbf{\Phi}$ and is parameterized by $\vartheta$. And the chosen action can be denoted as
\begin{equation}
a^{A} = \pi_{\vartheta}(\mathbf{\Phi})
\end{equation}
where we have $a^{A} \in \mathcal{A}$. Since the action space is discrete, we use softmax function at the output layer of the actor network so that we can obtain the scores of each action. The scores sum up to $1$ and can be regarded as the probabilities to obtain a good reward by choosing the corresponding actions.

\emph{Critic:} The critic is employed to estimate the value function $V(\mathbf{\Phi})$. At time instant $t$, when the action $a^{A}_t$ is chosen by the actor network, the agent will execute it in the environment and send the current observation $\mathbf{\Phi}_t$ along with the feedback from the environment to the critic. The feedback includes the reward $r_t$ and the updated observation $\mathbf{\Phi}_{t+1}$. Then, the critic calculates the TD error:
\begin{equation} \label{eq:TDerror}
\delta_{t} = r_t + \gamma V_{\mu}(\mathbf{\Phi}_{t+1}) - V_{\mu}(\mathbf{\Phi}_t)
\end{equation}
where $\gamma \in (0,1)$ is the discount factor.

\emph{Update:} The critic is updated by minimizing the least squares temporal difference (LSTD):
\begin{equation}
V^* = \arg \min_{V_{\mu}} (\delta_{t} )^2
\end{equation}
where $V^*$ denotes the optimal value function.

The actor is updated by policy gradient. Here, we use the TD error to compute the policy gradient\footnote{In (\ref{eq:policygradient}), policy gradient is denoted by $\nabla_{\vartheta} J(\vartheta)$ where $J(\vartheta)$ stands for the policy objective function, which is generally formulated as the statistical average of the reward.}:
\begin{equation}\label{eq:policygradient}
\nabla_{\vartheta} J(\vartheta) = E_{\pi_{\vartheta} } [ \nabla_{\vartheta} \log \pi_\vartheta(\mathbf{\Phi}, a^{A})  \delta_{t} ]
\end{equation}
where $\pi_\vartheta(\mathbf{\Phi}, a^{A})$ denotes the score of action $a^{A}$ under the current policy. Then, the weighted difference of parameters in the actor at time $t$ can be denoted as $\Delta\vartheta_{t} = \alpha \nabla_{\vartheta_t} \log \pi_{\vartheta_t}(\mathbf{\Phi}_t, a^{A}_t) \delta_{t}$, where $\alpha \in (0,1)$ is the learning rate. And the actor network can be updated using the gradient decent method:
\begin{equation}
\vartheta_{t+1} = \vartheta_t + \alpha \nabla_{\vartheta_t} \log \pi_{\vartheta_t}(\mathbf{\Phi}_t, a^{A}_t) \delta_{t}.
\end{equation}

\subsection{Operational Modes}
Once the DRL agent is initialized, it switches between two different modes: listening mode and attacking mode.

\begin{itemize}
	\item \emph{Listening mode:} In this mode, the DRL agent only observes the environment and updates its own policy based on the reward, but does not jam the selected channels so that the victim is not influenced and does not adopt a new policy.
	\item \emph{Attacking phase:} In this mode, the DRL agent jams the selected channels and decides whether to update its neural networks based on the victim's performance. When the victim performs well, the DRL agent should evolve its policy as the victim gradually adapts to the attacker's influence. However, when the victim performs poorly, the DRL agent should stop learning from the reward. Because in this situation the victim may frequently choose channels in bad states, and the reward may misguide the attacker.
\end{itemize}

We assume that the victim's model is pre-trained so that the victim's activity pattern is stable when the attacker starts to train its own neural networks. In this training phase, the DRL attacking agent works in the listening mode. And when the DRL agent is well trained, it can start the dynamic attack which we describe in detail in the following subsection.

\subsection{Dynamic Attack}
\begin{figure}
	\centering
	\includegraphics[width=1.\linewidth]{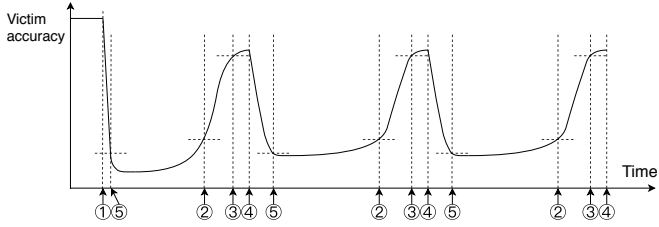}
	\caption{Stop-retrain-attack procedure of dynamic attack:
		\textcircled{\raisebox{-0.9pt}{1}} initial attack
		\textcircled{\raisebox{-0.9pt}{2}} stop attack
		\textcircled{\raisebox{-0.9pt}{3}} start retraining
		\textcircled{\raisebox{-0.9pt}{4}} stop retraining and start attack
		\textcircled{\raisebox{-0.9pt}{5}} stop updating the model}
	\label{fig:RAS}
\end{figure}

The DRL attacker uses the stop-retrain-attack (SRA) procedure shown in Fig. \ref{fig:RAS}. DRL attacker aims at avoiding the situation in which the victim learns a totally new action policy once the model is well trained. For this purpose, the duration of each cycle of the DRL attacker is fixed at a certain value that prevents the victim to update to a new policy.

As shown in Fig. \ref{fig:RAS}, the DRL attacker starts its first attack at time \textcircled{\raisebox{-0.9pt}{1}} when the victim model has been working in a stable fashion and working well. Before this point, the DRL attacker works in listening phase to learn the victim's activity pattern, and we assume that at time \textcircled{\raisebox{-0.9pt}{1}}, the DRL attacker can also function well with high stability. Once the attack is initiated, the performance of the victim drops rapidly. In this process, the victim keeps updating its model to overcome the influence of the attacks, and at the same time, the attacker also keeps updating its model to adapt to the victim's changing policy. However, we should note that the attacker is always encouraged to choose the same channel as the victim does. Hence, when the victim is forced to explore other channels which are not attacked in order to find a new policy to counteract against the attacks, it cannot avoid but try bad channels in order to find the good ones. From the perspective of DRL attacker, there is no need to follow the victim's selection because the victim's model updates dramatically and the policy may perform worse initially. On the one hand, it is difficult for the attacker to learn an unstable policy. On the other hand, copying the bad policy may give victim the chance to recover its performance. Based on this idea, the DRL attacker stops updating when the performance of the victim is lower than a threshold and we mark this time instant as time \textcircled{\raisebox{-0.9pt}{5}}. Though the DRL attacker model stops learning, it still works in attacking mode, so the performance of the victim continues to decrease. As mentioned before, the DRL attacker should stop jamming the channels before the victim adapts to its attacks, because the victim is also a reinforcement learning agent that has the ability to act against attacks naturally. At time \textcircled{\raisebox{-0.9pt}{2}}, the victim's performance starts to recover, meaning that a new policy is being formed in the victim model. To avoid pushing the victim to the new policy further, the DRL attacker needs to switch to the listening mode at time \textcircled{\raisebox{-0.9pt}{2}} to encourage the victim to return to its old policy as quickly as possible. And at time \textcircled{\raisebox{-0.9pt}{3}}, the victim is able to perform as well as that before the attack, and the DRL attacker will start to retrain its model and keep working in listening mode to adjust its policy based on the victim's activity until time \textcircled{\raisebox{-0.9pt}{4}} when the DRL attacker switches to attacking mode and starts a new cycle. In our implementation, the duration of each cycle is fixed to $2000$ time slots, and the gap between time \textcircled{\raisebox{-0.9pt}{3}} and \textcircled{\raisebox{-0.9pt}{4}} is fixed at $200$ time slots. Also, in the experiments, the duration between time \textcircled{\raisebox{-0.9pt}{2}} and \textcircled{\raisebox{-0.9pt}{3}} is very small.

\subsection{Experiments}\label{subsec: RL exp}
In this section, we test the proposed DRL attacker with a well-trained victim model and channel pattern introduced in Section \ref{Sec: pre}.

\begin{figure}
	\centering
	\includegraphics[width=0.7\linewidth]{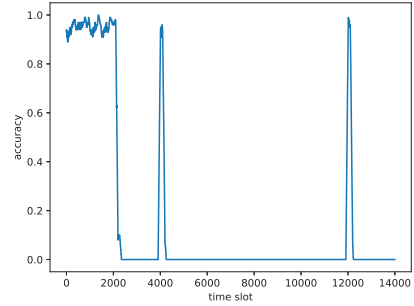}
	\caption{Victim's accuracy under DRL attacker's SRA procedure. The victim does not employ an $\epsilon$-greedy policy (i.e., $\epsilon = 0$).}
	\label{fig:rlattackerepsilonv}
\end{figure}

First, we test the DRL attacker under the condition that victim model works with $\epsilon = 0$ to show its full power. In Fig. \ref{fig:rlattackerepsilonv}, we plot the victim's accuracy over time to show the attackers' performance. The victim's policy crashes immediately after the DRL attacker starts jamming the channels at time slot $t = 2000$. However, the victim's policy can recover only for a short period of time after a few thousands of time slots. These brief recoveries are due to the fact that, as a reinforcement learning-based agent, the DRL attacker operates with an $\epsilon$-greedy policy with $\epsilon = 0.1$. The randomness in the DRL attacker's policy leads to a small chance for the victim to recover its performance from time to time.

\begin{figure}
	\centering
	\includegraphics[width=0.7\linewidth]{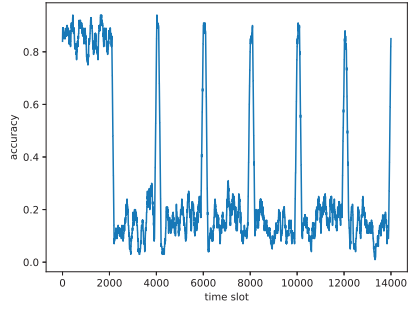}
	\caption{Victim's accuracy under DRL attacker's SRA procedure. The victim employs $\epsilon$-greedy policy with $\epsilon = 0.1$.}
	\label{fig:rlattackerrras}
\end{figure}

We note that it is not challenging for the proposed to attacker to jam the channels selected by the victim most of the time, and considering this, we consider the more adaptive victim model with $\epsilon = 0.1$ in the following experiments to show the performance of the proposed attacker facing a stronger victim user. In Fig. \ref{fig:rlattackerrras}, we plot the accuracy of the stronger victim with $\epsilon = 0.1$ under DRL attacker's SRA procedure. The DRL attacker stops updating the policy when the victim's accuracy is lower than 30\% and switches to the listening mode when the victim's accuracy recovers to higher than 30\% or if the duration of the current cycle is longer that $2000$ time slots. In the listening mode, the DRL attacker reloads its initial policy and retrains for 200 time slots before the next attacking mode starts. In Fig. \ref{fig:rlattackerrras}, we observe that the DRL attacker is able to have the victim's performance drop substantially and the recovery occurs over a very short period of time but the performance drops again significantly, which means that the victim operates with very low accuracy most of the time. Hence, under the DRL attacker's SRA procedure, the victim's accuracy is effectively limited to a low level. We also note that the pattern in Fig. \ref{fig:rlattackerrras} is similar to the one predicted in Fig. \ref{fig:RAS}.

\begin{figure}
	\centering
	\includegraphics[width=0.7\linewidth]{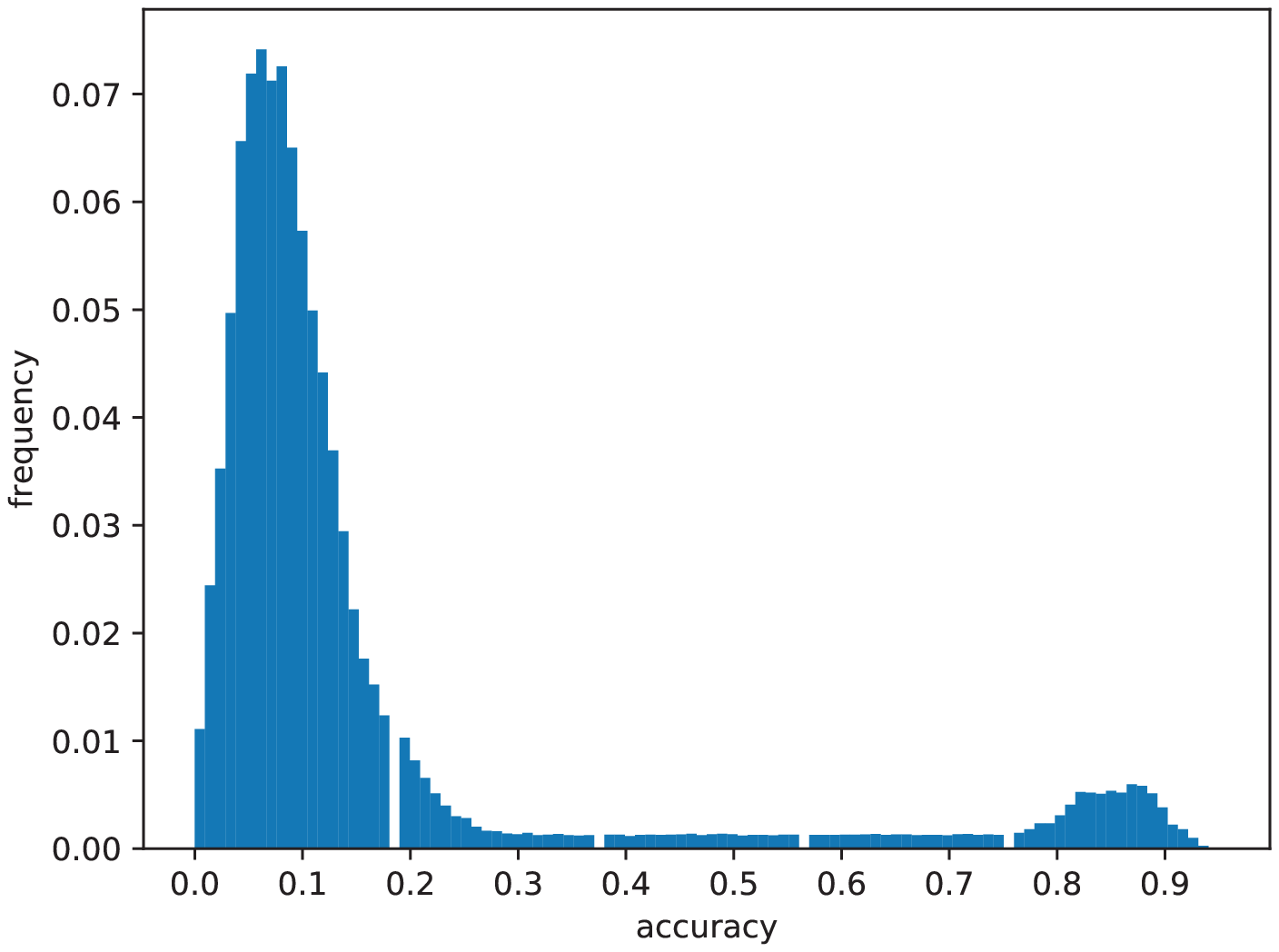}
	\vspace{-0.2cm}
	\caption{Empirical PDF of victim's accuracy under DRL attacker's SRA procedure.}
	\label{fig:rlattackerpdf}
\end{figure}

\begin{figure}
	\centering
	\includegraphics[width=0.7\linewidth]{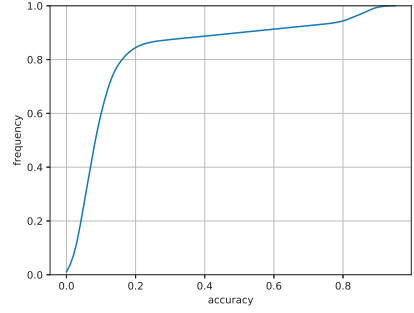}
	\vspace{-0.2cm}	
	\caption{Empirical CDF of victim's accuracy under DRL attacker's SRA procedure.}
	\label{fig:rlattackercdf}
\end{figure}

Next, we plot the corresponding empirical probability density function (PDF) and empirical cumulative distribution function (CDF) of the moving average of victim's accuracy in Figs. \ref{fig:rlattackerpdf} and \ref{fig:rlattackercdf} based on the accuracy results. Note that, to obtain the empirical PDF and CDF, we start collecting the accuracy data following the initial attack phase at time $t = 2000$. In Fig. \ref{fig:rlattackerpdf}, we observe that the victim's accuracy is highly concentrated around the level of $0.1$, and the corresponding empirical CDF in Fig. \ref{fig:rlattackercdf} exceeds 80\% when the accuracy is around 0.2. These results further indicate that the victim's accuracy is very low in the presence of jamming attacks. Specifically, the victim achieves accuracy levels of 0.2 or less around 80\% of the time.

\subsection{Impact of Limits on Attacker Power Budget}\label{subsec: budget exp}
In the SRA process, DRL attacker is assumed to always perform jamming attacks during the attacking phase, and stop during the listening phase periodically. To control the jamming power consumption, we in this section consider a budget on how frequently the attacker can perform jamming during the attack stage. In each time slot, the actor of attacker's actor-critic DRL agent gives an array of probabilities that the attacker should attack each of the channels, and the attacker attacks the channel with the highest probability $p_{attack, max}(t)$. Due to the limited power budget, the attacker now only attacks when $p_{attack, max}(t)$ is higher than a threshold $p_\theta$, otherwise it pauses the attacks and listens. Within each period of $T$ time slots, if the attacker has already attacked more than $\lfloor T\theta \rfloor$ times, it ceases the attacks until the next period so that the average jamming power consumption is limited by $\theta$ fraction of the peak jamming power, where $0<\theta<1$. At the end of the current period, the attacker updates the threshold $p_\theta$ with the recorded probabilities $\{p_{attack, max}(t), \text{ for } t = t_0, t_0+1, \ldots, t_0+T-1\}$. We rearrange the $T$ probabilities in descending order, and pick the $\lfloor T\theta_u \rfloor$th lowest probability as the updated threshold $p_\theta$, where, for instance, we can set $\theta_u=\theta$.

It is obvious that the actual ratio of attacked time slots is less than or equal to $\theta$, because the distribution of $p_{attack, max}(t)$ fluctuates over time, but $p_\theta$ is determined by the probabilities of the last period. If the probabilities in the current time period grow higher, the attacker tends to select more time slots to attack, and reach the threshold $\lfloor T\theta \rfloor$ before $t=t_0+T-1$, and then stop until the next period. In such a period, the attack ratio is $\theta$. On the contrary, if the probabilities drop lower, the attacker tends to choose less time slots to attack. Due to this, it will potentially not reach the threshold $\lfloor T\theta \rfloor$, and the attack ratio is less than $\theta$. Both scenarios occur in our simulations, and therefore the average attack ratio generally becomes less than $\theta$. Due to this, one may consider setting $\theta_u>\theta$, so that the first case (higher probabilities) is experienced more frequently and the second case less frequently. With this, the attack ratio increases. However, we have experienced that the victim accuracy also increases, which means we attack more frequently, but the attacker is weaker than the original setting $\theta_u=\theta$. It turns out that this modified attack policy tends to perform unnecessary attacks and the pattern is more predictable and hence less difficult to learn by the victim. Due to this, in the following experiments, we use $\theta_u=\theta$, and call this attack power budget parameter $\theta$ as the attack rate since this is also a measure of how frequently jamming attacks are performed within a given period.

\begin{figure}
	\centering
	\includegraphics[width=0.7\linewidth]{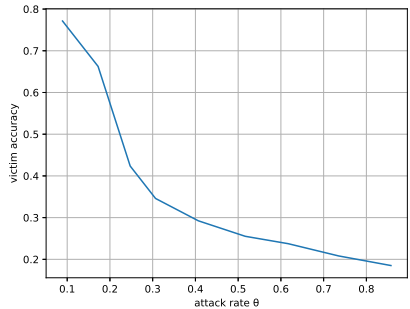}
	\caption{Victim's accuracy vs. attack ratio}
	\label{fig:acc_att_rate}
\end{figure}

\begin{figure}
	\centering
	\includegraphics[width=0.7\linewidth]{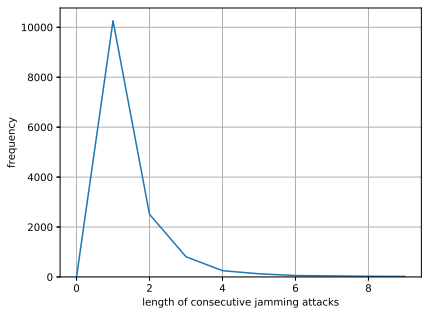}
	\caption{Empirical distribution of the length of consecutive jamming attacks}
	\label{fig:noise_length}
\end{figure}

In Fig. \ref{fig:acc_att_rate}, we plot the victim's accuracy as a function of the attack rate $\theta$ with $T=1000$. Different values of $T$ lead to similar curves. In this figure, we notice the elbow point at $\theta = 0.3$. Increasing the attack rate beyond this limit will not lead to a significant drop in the victim's accuracy. Therefore, $\theta = 0.3$ is a relatively reasonable budget on the attack rate, at which victim's accuracy is still suppressed to $34.6\%$. Fig. \ref{fig:noise_length} shows the empirical distribution of different lengths of consecutive jamming attacks. All jamming attacks are less than 10 time slots long (indicating that we have consecutive jamming attacks over at most 10 time slots before the attacker agent stops), and the peak is experienced at 1, so the jamming attack emission is intermittent and steady.

Above, we see that even the budgeted attack performs well in the absence of defensive strategies. In the following sections, we remove the budget constraint from the attacker agent and propose and test different defense strategies against the full power attacker, which represents the worst case scenario from the victim's perspective.

\section{Diversified Defense with PID Control}\label{sec: pid}
As shown in the last section, a jamming attack will lead to a significant drop in victim's accuracy. In the following three sections, we assume that the observed sudden decrease in accuracy is due to an attack, and we will address detection of attacks in subsection \ref{subsec: detection}. In this section, we introduce a diversified defense strategy that utilizes proportional-integral-derivative (PID) control based on recorded victim accuracy history, so that this strategy is universally applicable to any other type of adversarial jamming attacks.

\subsection{Diversified Defense}\label{subsec: div}
Aforementioned attacker agent can choose one channel at a time to perform the jamming attack, block victim’s transmission and significantly interrupt its learning process. In the presence of such attacks, we have observed in Section \ref{sec: RL} that the victim without any defense will have significantly lower accuracy. As the attacker continuously learns from interactions with the environment and victim, traditional defense methods \cite{yuan2019adversarial} do not work against such an attacker. If the victim attempts to predict which channel will be attacked and chooses a less wanted one on purpose, it will fall into a dilemma: the less the victim user deviates from the DRL policy, the weaker defense we have; the more it deviates from the DRL policy, the more likely that the victim model will suffer due to frequent selection of channels in bad state.

One alternative approach is to choose one channel from multiple options according to a certain probability distribution \cite{balcan2018diversified}. The actor of victim actor-critic DRL agent provides an array of predicted probabilities $p_{victim}$ of each channel being in good state. The original setting is to choose the channel with the highest $p_{victim}$, but now we rank them in descending order of $p_{victim}$, and assign a fixed array of probabilities that we choose for each channel. (In the original scheme, the second and third highest $p_{victim}$ vary several orders of magnitude over time, and highest probability can get close to 1, making the channel selection more predictable.) That is to say, for the channels $i = 1, 2, \ldots, N$ listed in the descending order of the original probabilities determined by the DRL agent (which may correspond to different channels at different times), we assign a fixed probability array $\{p(1), p(2), \ldots, p(N) \}$, where $\sum_{i = 1}^{N} p(i) = 1$ and $p(i)$ decreases as $i$ increases. Then the probability that the victim user chooses channel $i$ is
\begin{equation}\label{equ: constdiv}
P\{a_{victim}(t) = i\} = p(i).
\end{equation}

However, such a strategy fails to utilize the victim accuracy and has limited performance. To address this, we propose an additional PID control as discussed next.

\subsection{PID Control}\label{subsec: pid}
PID controller is a feedback control system \cite{aastrom2006advanced}. The error $e(t)$ is defined as the difference between the desired goal and measured position. PID controller calculates the weighted sum of proportional, integral, and derivative terms of $e(t)$ to adjust the control variable $u(t)$ as shown below:
\begin{equation}\label{equ: pid_general}
u(t)=K_p e(t)+K_i \int_{0}^{t}e(t')dt'+K_d \frac{de(t)}{dt}
\end{equation}
where $K_p$, $K_i$, $K_d$ are constant weights.

This approach is widely used to optimize automatic control since it can effectively correct systematic discrepancies and dampen oscillations. However, if we simply use the strategy with probabilities selected as in (\ref{equ: constdiv}), the probability to choose a channel will not change, regardless of how high or low the channel access accuracy becomes. Typically, as discussed in Section \ref{sec: RL}, dynamic attacker stops attacking and learns from the recovered victim periodically. Thus, when the accuracy abnormally increases, we also desire to suppress it by increasing the probability of accessing channels that are potentially in bad states. We apply the idea of PID control, and calculate the short term average, long term average and long term difference of victim channel access accuracy. We add these modified PID components to the probability array to dampen accuracy oscillations. For channels $i = 1, 2, \ldots, N$ listed in the descending order of $p_{victim}$, and for given victim channel access accuracy $\alpha(t)$, and constant arrays $p(i)$, $K_p(i)$, $K_i(i)$, $K_d(i)$, we modify (\ref{equ: constdiv}) as follows:
\begin{equation}\label{equ: pid3}
P\{a_{victim}(t) = i\} = P\{i\}(t) = \frac{P'\{i\}(t)}{\sum_{i'=1}^{N}P'\{i'\}(t)}
\end{equation}
where
\begin{equation}\label{equ: pid2}
P'\{i\}(t)=\max\{P''\{i\}(t), 0\} \quad \text{and}
\end{equation}
\begin{equation}\label{equ: pid1}
\begin{aligned}
P''\{i\}(t)= & \ p(i)+K_p(i)\sum_{t'=t-10}^{t-1}\alpha(t')+K_i(i)\sum_{t'=t-200}^{t-1}\alpha(t') \\
& +K_d(i)\left(\sum_{t'=t-200}^{t-1}\alpha(t')-\sum_{t'=t-400}^{t-201}\alpha(t')\right).
\end{aligned}
\end{equation}

We have $K_p(i)$, $K_i(i)$, $K_d(i)$ increase as $i$ increases. When the accuracy is low, this will boost the probability to choose channels predicted to be in good states and suppress the probability of choosing channels in potentially bad states, and vice versa. Thus, we improve the performance of the victim, while keeping the attacker confused at the same time.

\section{Diversified Defense with an Imitation Attacker}\label{sec: dummy}
In this section, we propose another diversified defense strategy which assumes that the attacker operates using an actor-critic DRL agent. As the victim has access to all the time series information with which the attacker trains its agent, it can train its own imitation actor-critic DRL attacker, which may have different neural network parameters and reward assignment from the true attacker. However, as both of them share the same goal to choose the victim's channel, the policy of the imitation attacker is expected to converge to that of the true attacker. This property of deep learning is called transfer-ability \cite{xie2019improving}.

Simply switching to the channel with the second highest probability when the imitation attacker provides a prediction of possible channels to be jammed is not a very effective strategy, because the dynamic DRL attacker can still learn from this pattern. Thus, we introduce another diversified defense to confuse the attacker with help from the imitation attacker by modifying (\ref{equ: constdiv}). Given that $i$ is the channel index rearranged in descending order of the probabilities provided by the victim DRL actor (considering which the original victim user chooses the first one, i.e., the channel with index $i =1$) and given that the imitation attacker chooses channel $a_{imitation}$, the probability that victim chooses channel $i$ is given as
\begin{align}\label{equ: dummy}
P\{a_{victim}(t) = i\} = \begin{cases}
\mathbf{1}(i = 1) \hspace{.3cm} \text{if $a_{imitation}(t) \neq 1$}\\
p(i) \hspace{.3cm} \text{if $a_{imitation}(t) = 1$}\\
\end{cases}
\end{align}
where $\{p(i)\}$ is a fixed set of probabilities (generally selected in a way to satisfy $p(1)<.5$ and  $p(2)>p(3)>p(4)>\ldots>p(N)$) and $\mathbf{1}(\cdot)$ is the indicator function.

Within this setting, when the true attacker has an accurate prediction, the victim gets the same information from its imitation attacker and assigns larger probabilities to channels that had originally lower probabilities assigned by the DRL channel access agent, and the attacker can barely learn from this diversified selection. When the attack model collapses and fails to attack, the victim simply relies on its prediction made by the DRL multichannel access agent, and is almost unaffected by the attack.

\section{Defense via Orthogonal Policies}\label{sec: 2policies}
In this section, we introduce a defense strategy with orthogonal policies based on transition matrices without having any assumptions on the attacker and environment. Given action space $\mathcal{A}$, if a policy chooses action $a_1$ at time $t$ and action $a_2$ at time $t+\tau$ and receives positive reward for both actions, we define this transition from $a_1$ to $a_2$ as a successful $\tau$th order transition. We collect all such transitions for orders $\tau=1,2,3,\ldots,T$ during a certain observation time period, and add up to a set of transition frequency matrices of size $N\times N$, where $N$ is the total number of actions, and the component at row $a_1$ and column $a_2$ in the matrix corresponding to the $\tau$th order transitions describes the number of $\tau$th order transitions from action $a_1$ to $a_2$. We then define the normalized correlation of two transition matrices $M_{\pi_1\tau}$, $M_{\pi_2\tau}$ of the same order $\tau$ from different policies $\pi_1$, $\pi_2$ as
\begin{equation}\label{equ: correlation}
R(\pi_1,\pi_2,\tau)=\frac{M_{\pi_1\tau}\circ M_{\pi_2\tau}}
{\sum\limits_{a_1 \in \mathcal{A}}{\sum\limits_{a_2 \in \mathcal{A}}{M_{\pi_1\tau}[a_1,a_2]}}
		\times \sum\limits_{a_1 \in \mathcal{A}}{\sum\limits_{a_2 \in \mathcal{A}}{M_{\pi_2\tau}[a_1,a_2]}}}
\end{equation}
where $\circ$ denotes element-wise Hadamard product, and $M_{\pi\tau}[a_1,a_2]$ denotes the element at row $a_1$ and column $a_2$ in the matrix corresponding to the $\tau$th order transitions of policy $\pi$. We define orthogonal policies as policies that have negligible normalized correlation. Since the orthogonal policies almost share no common action transitions, they have the longest distance in the action transition space. Therefore, once the attacker adapts to one of the orthogonal policies, it is difficult for the attacker agent to adapt to other orthogonal policies. Taking advantage of this, we minimize the normalized correlation to train a set of orthogonal policies. As an effective defensive strategy, the victim agent can switch between these orthogonal policies when being attacked.

\begin{figure}
	\centering
	\includegraphics[width=1\linewidth]{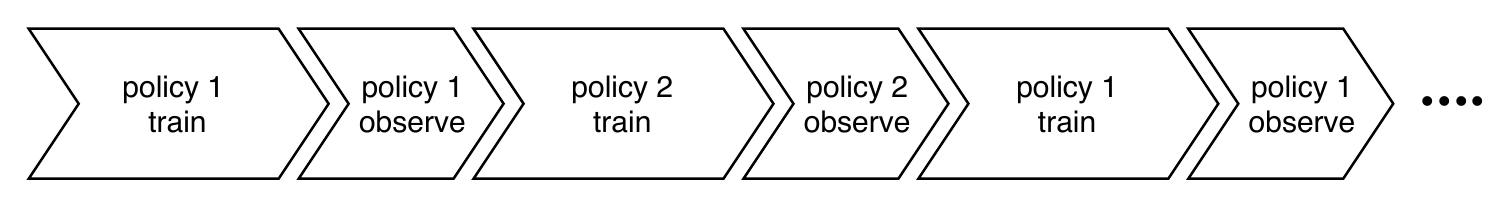}
	\caption{Training process for constructing two orthogonal policies}
	\label{fig:orthotrain}
\end{figure}

To apply this idea in the multichannel access defense problem, we design an iterative training process to train two orthogonal policies as depicted in Fig. \ref{fig:orthotrain}. We start with two copies of the trained victim model (as described in Section \ref{Sec: pre}), namely policy 1 and policy 2, and train them to minimize the normalized correlation between policy 1 and 2 alternately, so that each policy gradually deviates from each other, and becomes orthogonal. In each period, for example, we train policy 1 with high exploration probability $\epsilon=0.9$ to force it to explore a random channel with designed probability distribution $P(t)$, and then set $\epsilon=0$ for a short duration in listening mode to observe this trained policy and record its transition matrices $M_{\pi_1\tau}, \tau=1,2,3,\ldots,T$.

The probability distribution $P(t)$ is generated from policy 2 transition matrices $M_{\pi_2\tau}, \tau=1,2,3,\ldots,T$ observed in the last period. If policy 1 leads to the choice of channel $a(\tau)$ at time $t-\tau$, we extract the probability array as

\begin{equation}\label{equ: orthoprobp}
p(\tau)=\text{normalize}\left(\frac{1}{M_{\pi_2\tau}[a(\tau),:]+1}\right)
\end{equation}
where $M_{\pi_2\tau}[a(\tau),:]$ is the $a(\tau)$th row of $\tau$th order transition matrix for policy 2, and the normalization is introduced so that the probabilities are non-negative and sum up to one (as in (\ref{equ: pid3}) and (\ref{equ: pid2})). Therefore, we have the following weighted sum of such probability distributions over different $\tau$s:

\begin{equation}\label{equ: orthoprobP}
P(t)=\text{normalize}\left({\sum_{\tau=1}^{T}{\varrho^{\tau-1}\eta_\tau \left(p(\tau)-\frac{1}{N}\right)}+\frac{1}{N}}\right)
\end{equation}
where $N$ is the total number of channels, $0<\varrho<1$ is decay factor, and $\eta_{\tau}$ is the indicator whether agent received positive reward at time $t-\tau$:

\begin{align}\label{equ: theta}
\eta_\tau = \begin{cases}
+1.5 \hspace{.3cm} \text{if $r(t-\tau)=1$}\\
-0.5 \hspace{.3cm} \text{if $r(t-\tau)=-1$}.\\
\end{cases}
\end{align}

This distribution guarantees transitions that often appear in policy 2 are less explored by policy 1, which leads to orthogonality. Among these intensively explored transitions, DRL agent will only learn these successful actions, so the trained policies will be intermediate between $p(\tau)$ and the underlying environment, and maintain an acceptable accuracy.

We should be aware of one potential risk of this training strategy, that two policies might tend to choose two non-overlapping subset of channels. Such pair of policies has low normalized correlation, but the mixed transition pattern is easy to learn by the attacker, i.e., an averaged transition map works perfect as an attack whichever policy is picked. Therefore, we introduce a regularization term to (\ref{equ: orthoprobP}). When we train one policy, for example policy 1, we also record how many successful actions were performed during the last policy 2 observation period in a 1-dimensional array $p_{reg}$. With this, we have the regularized distribution

\begin{equation}\label{equ: orthoprobPreg}
P(t)=\text{normalize}\left({\sum_{\tau=1}^{T}{\varrho^{\tau-1}\eta_\tau \left(p(\tau)-\frac{1}{N}\right)}+\frac{1}{N}+\beta p_{reg}}\right)
\end{equation}
where $0<\beta<1$ is a weight factor. We intend, but not force, to lead the two policies to choose the same subset of channels, so that the mixed transition pattern is much harder to learn by the attacker. An additional benefit is that we can use a higher learning rate to train the policies. An actor-critic agent might collapse by choosing one channel forever without exploring other channels when the learning rate is too high, but the regularization term leads it to explore a chosen subset of channels by the other policy, and avoid such a collapse. Therefore, we are able to use higher learning rates to train faster.

\section{Experiments with Defensive Strategies}\label{sec: exp}
In this section, we test the proposed defense strategies against the well-trained dynamic attacker described in section \ref{sec: RL}. Before the experiments start, the victim is trained over 500000 time slots and the attacker has another 200000 time slots until convergence. The trained victim starts at $t=0$, and the trained attacker initiates attacking at $t=2000$. As victim accuracy drops dramatically, defense agents start at $t=2400$. In all diversified defense cases, we set $\epsilon=0$ (the probability to choose a channel randomly) at the victim DRL agent, as we have already deployed better randomized strategies.

\subsection{Diversified PID Defense}\label{subsec: pid exp}
\begin{figure}
	\centering
	\includegraphics[width=0.7\linewidth]{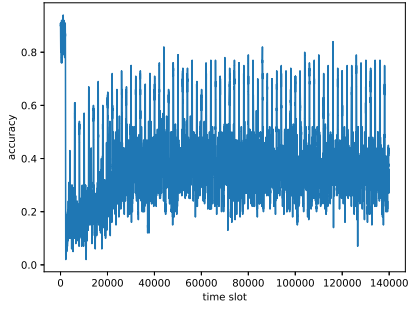}
	\caption{Victim's accuracy with diversified defense involving PID control.}
	\label{fig: pid}
\end{figure}

First, we test the diversified PID defense described in Section \ref{sec: pid}. We plot the moving average of the victim's accuracy in Fig. \ref{fig: pid}. During the initial 2000 time slots, the victim without the attacker performs well, but when the attack starts, the accuracy drops below $20\%$. As observed in Fig. \ref{fig:rlattackerrras}, the victim without any defense will have an average accuracy of $14\%$. Shortly afterwards, PID defense is initiated and the attacker model gradually collapses, so the victim accuracy rises to $39\%$, which is close to a half of the victim's performance in the absence of a jamming attacker. In this experiment, the underlying channel pattern has 2 out of 16 channels in good condition, and the attacker knows if the chosen channel is good and if the victim also chooses it after the attack. Thus, this result means that although the attacker learns from the victim's choice, we have managed to confuse the attacker to some extent.

\begin{figure}
	\centering
	\includegraphics[width=0.7\linewidth]{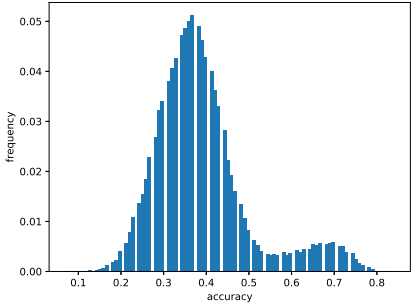}
	\caption{Empirical PDF of victim's accuracy under diversified PID defense.}
	\label{fig: pid_pdf}
\end{figure}

\begin{figure}
	\centering
	\includegraphics[width=0.7\linewidth]{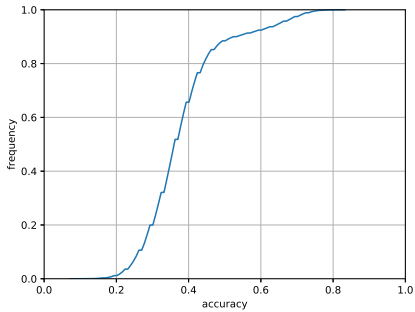}
	\caption{Empirical CDF of victim's accuracy under diversified PID defense.}
	\label{fig: pid_cdf}
\end{figure}

Figs. \ref{fig: pid_pdf} and \ref{fig: pid_cdf} show the corresponding empirical PDF and CDF of PID converged performance. Note that we have made an additional assumption only for the PID case that after being attacked, the victim knows if the chosen channel was good before attack. In this case, we set the victim reward as 0.5 (to indicate a partially good choice) instead of -1, encouraging the victim to try the underlying good channels. As PID assigns more randomness to channel selection, it is necessary to lead the victim towards good channels even if they have been attacked, to avert victim model from collapsing. The converged accuracy without such an assumption is about $30\%$.

\subsection{Diversified Imitation Defense}\label{subsec: dummy exp}

Next we test the diversified imitation defense described in Section \ref{sec: dummy}. We start to train the imitation attacker before the attack starts as well as the true attacker, but the reward setting is different. Given that the attacker chooses channel $a_{attacker}$, the victim chooses channel $a_{victim}$, and the underlying good channels are denoted by $G=\{i_1, i_2\}$, the reward of true attacker $r_T(t)$ is
\begin{align} \label{equ: att rew}
r_T(t) = \begin{cases}
+ 1 \hspace{.55cm} \text{if $a_{attacker} = a_{victim}$ and $a_{victim}\in G$,}\\
+ 0.5 \hspace{.3cm} \text{if $a_{attacker} = a_{victim}$ and $a_{victim}\notin G$,}\\
- 0.5 \hspace{.3cm} \text{if $a_{attacker} \neq a_{victim}$ and $a_{victim}\notin G$,}\\
- 1 \hspace{.55cm} \text{if $a_{attacker} \neq a_{victim}$ and $a_{victim}\in G$.}
\end{cases}
\end{align}

For the imitation attacker, we do not assume that the victim knows if an attacked channel was good or not, and therefore the imitation attacker only uses the criterion if it chooses the same channel as the victim does after the victim deploys diversified defense and switches to another channel. Thus, given the victim's channel selection $a_{victim}$ after the defense strategy is deployed, and the imitation selection $a_{imitation}$, the reward of the imitation attacker $r_I(t)$ is
\begin{align}
r_I(t) = \begin{cases}
+ 1 \hspace{.3cm} \text{if $a_{imitation} = a_{victim}$,}\\
- 1 \hspace{.3cm} \text{if $a_{imitation} \neq a_{victim}$.}\\
\end{cases}
\end{align}

\begin{figure}
	\centering
	\includegraphics[width=0.7\linewidth]{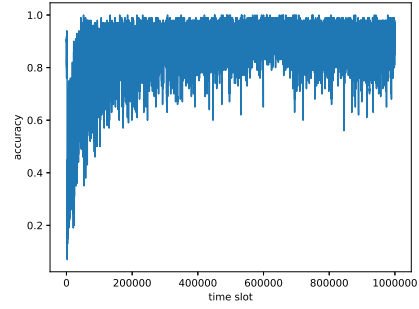}
	\caption{Victim's accuracy with diversified defense involving an imitation attacker.}
	\label{fig: dummy}
\end{figure}

We plot the moving average of the victim's accuracy in Fig. \ref{fig: dummy}. Initially, we have a similar performance as in the PID case, and after the diversified defense is initiated, the performance gradually converges to $90\%$ accuracy, indicating an outstanding defense against jamming attacks. Comparing with the performance without the attacker (shown in Fig. \ref{fig:victimaccuracybeforeattack}), the achieved accuracy is lower than the case of $\epsilon = 0$, where $95\%$ accuracy is attained, but higher than $85\%$ accuracy that is achieved with the $\epsilon$-greedy policy.

\begin{figure}
	\centering
	\includegraphics[width=0.7\linewidth]{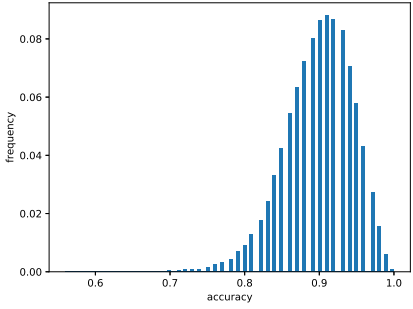}
	\caption{Empirical PDF of victim's accuracy under diversified imitation defense.}
	\label{fig: dummy_pdf}
\end{figure}

\begin{figure}
	\centering
	\includegraphics[width=0.7\linewidth]{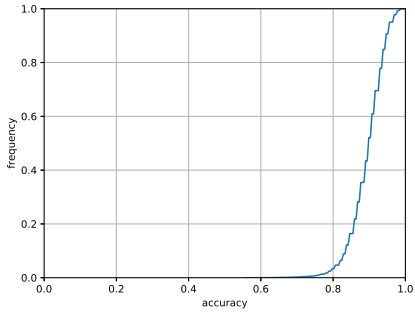}
	\caption{Empirical CDF of victim's accuracy under diversified imitation defense.}
	\label{fig: dummy_cdf}
\end{figure}

We also observe the significantly improved performance in accuracy in Figs. \ref{fig: dummy_pdf} and \ref{fig: dummy_cdf} where empirical PDF and CDF are provided, respectively. We additionally note that the number of layers and nodes in the imitation attacker's neural network does not affect the performance, as long as the imitation attacker has the purpose to select the same channel as the victim does. The reward assignment of the imitation attacker can also vary. For instance, if we set it the same as the true attacker as in (\ref{equ: att rew}) or simply use $[a_{victim}\in G]$, the difference in the final accuracy is less than $2\%$. With all different settings, imitation attacker converges to the true attacker, with $82\%$ probability on average in choosing the same channel. So, this defense strategy is robust against different settings, and successfully misleads the actor-critic dynamic attacker, rendering it a very effective strategy against such attacks.

\subsection{Defense via Orthogonal Policies}\label{subsec: 2policy exp}
\begin{figure}
	\centering
	\includegraphics[width=1\linewidth]{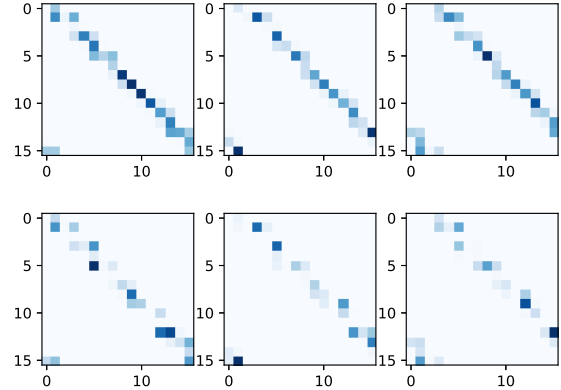}
	\caption{First row: the transition matrices for policy 1 with $\tau=1,2,3$. Second row: the transition matrices for policy 2 with $\tau=1,2,3$.}
	\label{fig: transmat}
\end{figure}

As another defensive mechanism, we introduce two orthogonal polices as discussed in Section \ref{sec: 2policies}. Note that these two policies are obtained via the victim neural networks. We retrain two copies of the victim's neural networks to minimize the normalized correlation of their transition matrices. In Fig. \ref{fig: transmat}, we show the first three pairs of transition matrices for two trained policies where $\epsilon=0$ (the probability to choose a random channel). We can see that policy 1 tends to select the channels that have adjacent indices, while policy 2 skips adjacent channels most of the time. In the absence of attacks, policy 1 has an accuracy of $80.5\%$ and policy 2 has an accuracy of $71.5\%$. (Note that we should only use initial victim model when there is no attack).

\begin{figure}
	\centering
	\includegraphics[width=0.7\linewidth]{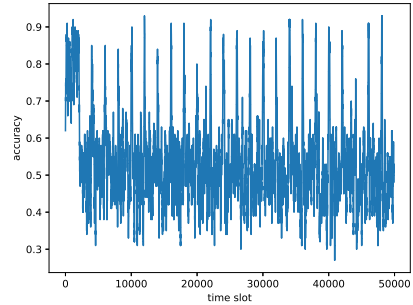}
	\caption{Victim's accuracy with defense via orthogonal policies.}
	\label{fig:orthogonalacc}
\end{figure}

\begin{figure}
	\centering
	\includegraphics[width=0.7\linewidth]{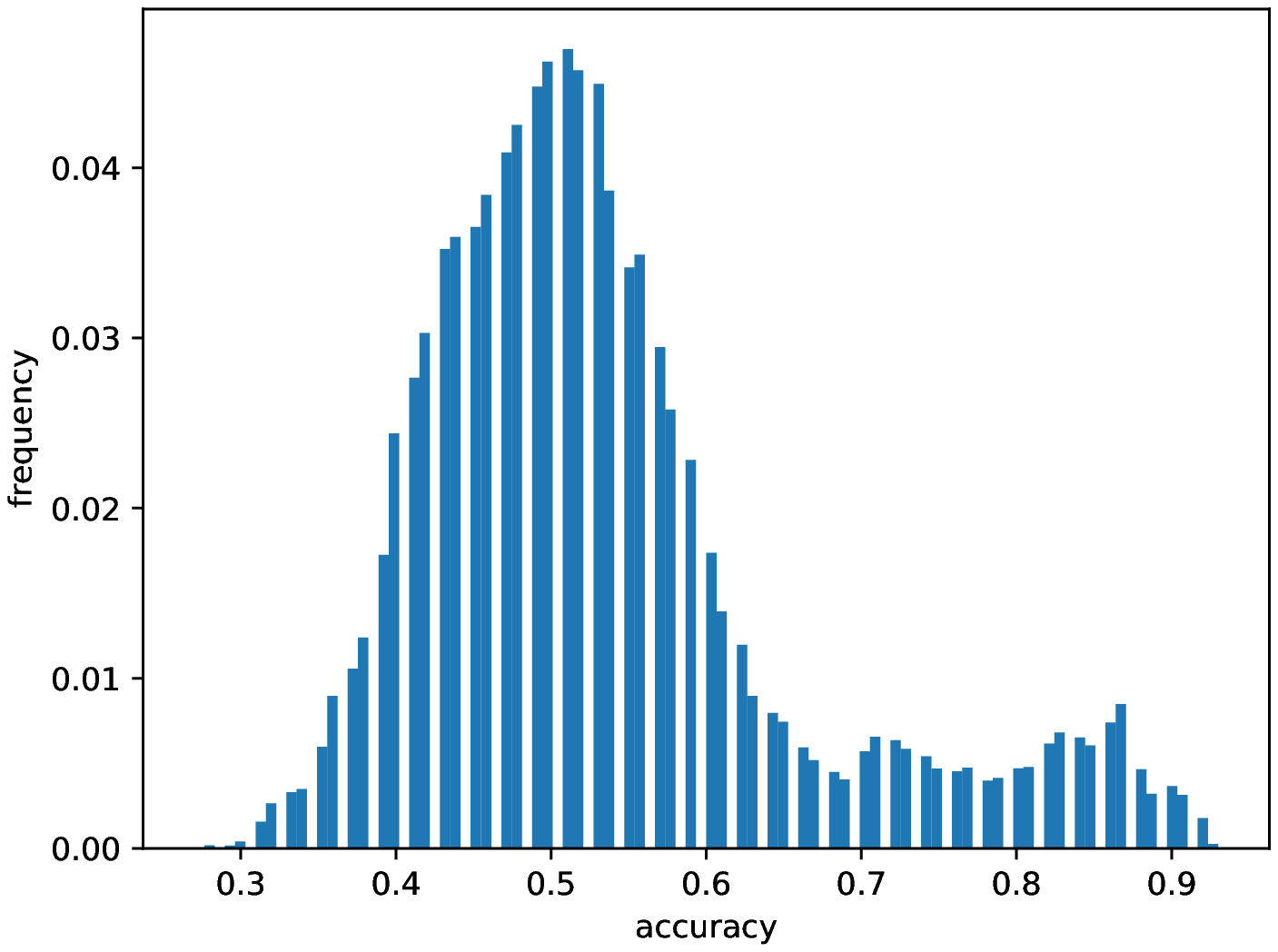}
	\caption{Empirical PDF of victim's accuracy under defense with orthogonal policies.}
	\label{fig:orthogonalpdf}
\end{figure}

\begin{figure}
	\centering
	\includegraphics[width=0.7\linewidth]{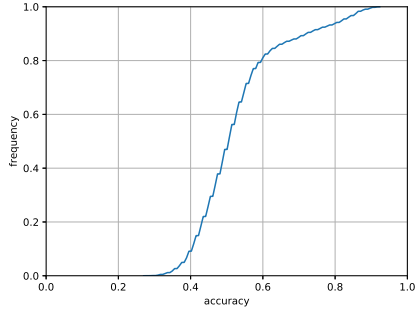}
	\caption{Empirical CDF of victim's accuracy under defense with orthogonal policies.}
	\label{fig:orthogonalcdf}
\end{figure}

In the defense phase, we reload these two policies iteratively to confuse the attacker, so that the attacker learns from one policy to the other all the time, and fails to perform a perfect attack. For every $2000$ time slots, we check if victim accuracy is lower than a certain threshold (e.g., $40\%$), and reload the other model if current policy fails. In Fig. \ref{fig:orthogonalacc}, we plot the victim accuracy when orthogonal policies are deployed as a defense strategy. The average accuracy increases from $14\%$ (before defense, as shown in Fig. \ref{fig:rlattackerrras}) to about $52\%$. Since the policy learned by the attacker is not orthogonal to the two designed orthogonal policies, it can be more similar to one of the two orthogonal policies (which is policy 1) than to the other policy (which is policy 2). However, since the two policies shown in Fig. \ref{fig: transmat} only share a few action transitions, the attacker's neural network cannot resume its high performance in a short period of time. With our strategy to switch policies as victim accuracy decreases, the overall performance almost remains the same over time. Figs. \ref{fig:orthogonalpdf} and \ref{fig:orthogonalcdf} show the corresponding empirical PDF and CDF of the accuracy.

\subsection{Attack Detection}\label{subsec: detection}
In the above evaluation of defense strategies, we have assumed that a sudden drop in victim's accuracy is always due to an adversarial attack. However, when the statistical description of the environment changes, the performance will also diminish. Since it will take a long time to retrain the user's DRL model, it is critical to distinguish environmental change from an adversarial jamming attack. Therefore, we propose an attack detection strategy based on orthogonal policies defense and diversified imitation defense.

\begin{figure}
	\centering
	\includegraphics[width=0.7\linewidth]{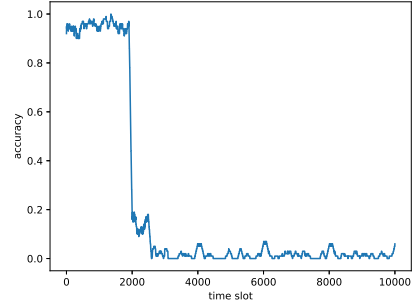}
	\caption{User's accuracy with orthogonal policies defense when environment changes}
	\label{fig:env_change_ortho}
\end{figure}

\begin{figure}
	\centering
	\includegraphics[width=0.7\linewidth]{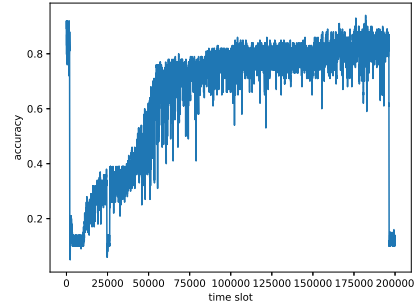}
	\caption{User's accuracy with diversified imitation defense when environment changes}
	\label{fig:env_change_dummy}
\end{figure}

We define the new environment as another round-robin switching pattern, where the order of 16 channel list is shuffled, and switch to this pattern occurs at $t=2000$. Figs. \ref{fig:env_change_ortho} and \ref{fig:env_change_dummy} show how the accuracies of the users employing the orthogonal policies and imitation attacker strategy, respectively, vary when the environment changes. On the one hand, for the user with imitation defense, the performance after an environmental change (as depicted in Fig. \ref{fig:env_change_dummy}) is not significantly different from that under a jamming attack (see e.g., Fig. \ref{fig: dummy}) as both converge towards $90\%$ accuracy. Therefore, imitation defense strategy does not work as a detector of attacks although it performs well at last. Additionally, in the case of an environment change, the accuracy recovers with more fluctuations and more slowly than the DRL agent without defense strategies because of the disturbance from the imitation attacker. On the other hand, for the user employing orthogonal policies, the accuracy almost goes to zero immediately when the environment changes (Fig. \ref{fig:env_change_ortho}), which strongly contrasts to Fig. \ref{fig:orthogonalacc}, where the accuracy under jamming attack fluctuates around $55\%$ as soon as the attack starts, with a minimum above $20\%$.

\begin{figure}
	\centering
	\includegraphics[width=0.7\linewidth]{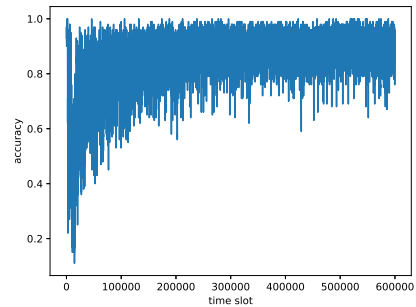}
	\caption{Victim's accuracy under attack with orthogonal policies defense as detector at $t=2000$, confirm the attack, and switch to diversified imitation defense at $t=10000$}
	\label{fig:detector}
\end{figure}

Thus, we propose to use orthogonal policies as an attack detection mechanism as follows. When the accuracy drops, we can first apply the defense with orthogonal policies. If the accuracy recovers and fluctuates with this approach, we confirm that it is an adversarial jamming attack, and switch to diversified imitation defense for better protection. The performance in this scenario is provided in Fig. \ref{fig:detector}. If, on the other hand, the accuracy does not recover even after orthogonal policies, we decide that the environment has changed and we retrain the DRL agent without initiating defense.

\section{Conclusion}\label{sec: con}
In this paper, we have introduced a dynamic actor-critic DRL jamming attacker aimed at minimizing the accuracy of a victim user performing dynamic multichannel access using its own DRL agent. We have introduced the actor-critic architecture of the proposed DRL attacker, and then presented its SRA working procedure. We have demonstrated that the DRL attacker can perform effectively in the absence of defense. In the performance evaluation of the DRL attacker, we have also considered a jamming power budget, limiting the attack rate. We have observed that with only $30\%$ power consumption (or equivalently attack rate), we can still suppress victim's accuracy to $34.6\%$. In these analyses, we have specifically conducted experiments with a stronger victim that applies the $\epsilon$-greedy policy.

We also introduced three different diversified defense strategies, namely PID control, imitation attacker and orthogonal policies, against this attacker to improve the accuracy of the victim DRL agent. We have investigated how to set the probabilities in choosing different channels to confuse the attacker. We note that the diversified PID defense neither makes any assumption on the attacker nor acquires extra training beforehand and in this setting leads the victim user to reach an accuracy level of $39\%$. The diversified defense based on the immitation attacker assumes that the attacker employs an actor-critic DRL agent, and can attain a very high accuracy ($\approx90\%$) and is robust to different settings. As to the defense via orthogonal policies, we need to train the policies before the attack with no assumptions, and we achieve an accuracy of $50\%$. Based on orthogonal policies, we have also introduced an attack detection strategy that can differentiate attacks from changes in the environment.

\bibliographystyle{IEEEtran}
\bibliography{ref_AA}

\end{document}